# Is the Past Determined?

HERVE ZWIRN

*CMLA (ENS Paris-Saclay, France) & IHPST (CNRS, France) & LIED (Paris 7 University)*

*herve.zwirn@gmail.com*

**Abstract:** In a recent paper [1], I argued against backward in time effects used by several authors to explain delayed choice experiments. I gave an explanation showing that there is no physical influence propagating from the present to the past and modifying the state of the system at a time previous to the measurement. However, though the solution is straightforward in the case of delayed choice experiments involving only one particle, it is subtler in the case of experiments involving two entangled particles because they give rise to EPR-like situations. Considering that a measurement is not an actual change of the physical state of a system and is relative to the observer allows to understand that there is neither backward in time effects nor instantaneous collapse of the second system when the first one is measured, as is often postulated. This allows also to get rid of any non-locality [2]. In this paper, I want to go further into the consequences of this way of considering the measurement, that I have called Convivial Solipsism, and show that even if, in the usual sense, there is no physical effect of the present or of the future on the past, we must nevertheless consider that the observer's past is sometimes not entirely determined and that it becomes determined only when certain measurements are done latter. This apparent contradiction disappears if one understand that each observer builds, through her own measurements, her own world (that I call the phenomenal world in Convivial Solipsism) which is different from what we are used to consider as the common world shared by everybody.

## 1. Introduction

The conviction that the past is fixed and determines the present and the future is one of the firmest ideas, deeply anchored not only in everybody's folk conception but also in the foundations of the scientific method. What happened in the past is perfectly determined, whether or not we know it, and we cannot change it. In physics the past state of a system is used to predict its future state through an evolution equation. Challenging this conviction is always considered as absurd. This is why Wheeler's delayed choice experiments, seeming to imply that the behavior of a particle at one time is influenced by something decided latter, have been so discussed and have given birth to many papers defending various ways to understand what happens in this kind of experiment. Often just used to emphasize the weird consequences of quantum mechanics[1], these experiments have sometimes led some authors to defend real backward in time effects of the present on the past [4] or advanced waves [5].

This paper is largely based on a previous paper [1] where I gave an analysis which shows that it is perfectly possible to explain the delayed choice experiments results without any physical backward in time effects. In this previous paper, I presented the results inside a standard realist stance about the physical world (the physical world is considered as a universal reality shared by everybody) and showed that there is no physical effect propagating backward in time. This is easy to prove for experiments involving only one particle. Indeed in this case the paradoxical conclusion of an effect from the future to the past is simply based on a rough mistake in the interpretation of the measurement of the impulsion of the particles. However, when considering experiments involving entangled particles, things are more complex because understanding these EPR-like kind of experiments raises the whole measurement problem and the questions of non-locality and of causality. My explanation used for that the interpretation of the measurement process adopted in Convivial Solipsism [1, 2, 6, 7, 8]. But actually, I presented this explanation in a kind of mixed framework where, for the sake of simplicity, I continued adopting a standard realist stance though using some concepts of Convivial Solipsism. However, Convivial Solipsism leads to a distinction between what I call the empirical world and the phenomenal world that I did not explain in this paper.

---

[1] See [3] where the authors explain that is seems that the present has influenced the past (*'we have a strange inversion of the normal order of time'*) and that the strange predictions of quantum mechanics are once more verified, without analysing further this sentence.

In the following, I will make clear the distinction between these two kinds of worlds and I will show that in the phenomenal world which is relative to the observer, it can happen that a past property A of a system $S_1$ at $T_{-1}$ is not yet determined at a time $T_0 > T_{-1}$, but becomes determined when the observer makes at $T_1 > T_0 > T_{-1}$ a measurement of another property B (possibly on another system $S_2$) whose result is linked to the value that A should have had at $T_{-1}$ to be coherent with the result obtained on B at $T_1$. However this does not mean that there is a back in time physical influence of the measurement of B made on $S_2$ at $T_1$ such that the physical state of $S_1$ at $T_{-1}$ changes.

To understand this seeming contradiction we need to give a presentation of Convivial Solipsism and of the difference between the empirical world and the phenomenal world. This is what I will do first. I have already presented Convivial Solipsism in a series of previous papers [1, 2, 6, 7, 8] and I refer the reader to these papers for a detailed analysis. But for the reader who wants to understand this paper without being obliged to read these previous papers, I will give first a survey of Convivial Solipsism and will mainly insist on the introduction of the concepts of phenomenal world and empirical world. I will then complete the explanation of the delayed choice quantum eraser that I gave in [1] and I will after that extend these results to past events in general.

The plan of the paper is the following:

- In part A, I present a summary of Convivial Solipsism - which is a new interpretation that I developed in order to solve the measurement problem - and I introduce precisely the difference between what I call the empirical reality and the phenomenal reality, which has not been done previously. I also apply it to give a better understanding of some strange features of the measurement of entangled systems, in particular in the EPR-like situations.

- In part B, in order for this paper to be self-sufficient, I start by briefly summarizing the main conclusions of [1] on the delayed choice experiments. I show that these experiments do not need to postulate any backward in time effects, but that they emphasize the difficulty to understand the measurement in the usual standard way, so that a new interpretation is needed if one wants to be fully coherent. That's one of the motivations for developing Convivial Solipsism. Then I complete the explanation of what happens in the two entangled particles delayed choice experiment (quantum eraser) from an observer's perspective which had not been considered in [1].

- In part C, I discuss the consequences of Convivial Solipsism on what an event is and the way we must consider past events.

## 2. Part A: A brief presentation of Convivial Solipsism (ConSol)

I have already given a description of Convivial Solipsism (ConSol) in several previous papers [1, 2, 6, 7, 8] and I refer the reader to these papers for a detailed analysis. However, in order for this article to be self-sufficient, I recall here the main features of ConSol.

### 2.1 The basic tenets (informal presentation)

As is well known, the measurement problem comes from the difficulty to define rigorously what a measurement is inside the standard quantum formalism. The Schrödinger equation has to be used when no measurement is made, otherwise it is the reduction postulate. But depending on the fact that we include or not the measurement apparatus (and possibly the observer) in the system considered, an ambiguity arises and it is possible to argue in favor of an entanglement of the system and the apparatus (if the apparatus and the system are considered as a single big system) or in favor of a reduction of the state of the system if the apparatus is left outside. No totally satisfying solution to this problem has been provided yet despite the thousands of pages that have been written about it. ConSol is the result of a trial to solve the measurement problem inside a non-modified quantum formalism (so in a different way from Bohm's theory [9] or Girardi-Rimini-Weber's theory [10] for example).

Long ago, for solving the measurement problem, Wigner [11], London and Bauer [12] assumed that a definite result is obtained only when an observer gives a look at the apparatus. They postulated that the observer's consciousness has a direct physical impact on the state of the system and is the cause of the collapse. But this dualist position is now abandoned. So, how and when is a definite result obtained? In the interpretations that do not modify the quantum formalism, the answer differs accordingly to what is assumed. Everett [13] assumes that the observer splits in as many different observers as there are possible results and that each one is conscious of one result. Hence there are many different worlds each one corresponding to one possible result. However, inside each world, the result is shared among the different observers living in this world. Independently of the issue of dealing with a continuous infinity of worlds, Everett's interpretation faces the problem that it is impossible to give a meaning to probabilities[2]. Relational interpretation [14] solves the measurement problem through a confusion between measurement and correlation and so is not acceptable for me. Things are more complex with QBism [15]. I refer the reader to the analysis of QBism I wrote [16] where I show that the question of understanding precisely the status of the result of a measurement inside QBism is not totally settled by the authors of this position. I have explained precisely elsewhere [7, 16] why I think that none of these previous attempts is successful and I have detailed the issues they face. I will here only give the solution that ConSol brings to the measurement problem.

One of the main tenets of ConSol is the way a measurement is described. As described more formally below in § 2.2.1, a measurement necessarily involves an observer and is defined as the awareness of a result by an observer whose perception selects at random (according to the Born rule) one branch of the entangled state vector of the system written in the preferred basis and hangs-on to it. This is called the hanging-on mechanism. The direct consequence of the hanging-on mechanism is that a measurement is necessarily relative to an observer since it is nothing else than the perception of a definite result (selected at random among the possible results) by an observer. A consequence of that is that, contrary to what is often assumed, a simple interaction between an apparatus and a system (such as for example a particle going through a Stern and Gerlach apparatus) does not provide a measurement since a measurement needs to involve an observer.

Let's be more precise: A measurement is not a physical action modifying the physical state of a system but is the perception by the observer of one of the possible results when the system is in a superposed state of eigenstates of the observable that is measured. Similarly to Everett's interpretation, there is no collapse of the wave-function. When she does a measurement, the observer is only able to perceive one of the components of the superposed state (or of the entangled state between the system and the apparatus). The observer can only see one branch of the entangled wave function[3]. The state of the system is not modified and remains superposed (and even entangled with the apparatus and possibly with other systems and the environment). But once an observer has perceived a definite result, she stays hung-on to the branch she perceived for all the subsequent measurements and she will be able to perceive only results that are related to components linked to this branch. This guarantees that repeating the same measurement immediately after a first one will produce the same result even if the physical state of the system has not been reduced.

Now we can introduce the definition of the empirical world and of the phenomenal world. The empirical world of each observer gathers all the potentialities that the observer could actualize. It is described by a global entangled wave function (which is often called the universal wave function in the Everett's interpretation) and evolves only through the Schrödinger equation. It remains entangled and no reduction happens. In a way, it is everything that is potentially accessible to the observer, all the potentialities that the observer could actualize. Furthermore it is relative to each observer[4].

---

[2] In spite of the many attempts to save the probabilities inside Everett's interpretation, it remains doubtful that any of them succeeds.
[3] See more detailed explanations of the reasons why the observer can see only one branch in [2, 7, 8].

Now, every time an observer makes a measurement she selects through her perception one of the possible results. Then she remains hung-on to the corresponding selected branch of the global wave function of the empirical world. The totality of these results constitutes for this particular observer her own reality. This is what I call her phenomenal world. So each observer builds her phenomenal world through the measurements she does inside her empirical world which is not modified by these measurements. The phenomenal world is a tree of sub branches of the global entangled wave function[5]. Of course, the phenomenal world, which is built from the empirical world which is relative to the observer, is itself a fortiori relative to the observer. That is also true for the state vectors and the observables. This point is similar to what is supposed in QBism. Everything that can be expressed is always relative to one unique observer.

Now it is important to notice that an observer can have no direct access to the perception of the others. The consequence is that the phenomenal world of each observer is private and cannot be shared between different observers. Actually, for each observer, the other observers have a status similar to any other physical systems and can be in superposed states. Indeed, an observer can see that there are other "special" physical systems who are similar to her. This is why she can speaks of them as other observers. But having no way to directly witness their internal perception, the only way to communicate with them is to exchange by a physical interaction through any appropriate physical means and this is equivalent to perform a measurement on them.

Assume for example that Alice knows that Bob measured the spin along Oz of a spin half particle in an initial state "+" along Ox. Before Alice asks Bob what he saw, from Alice's point of view Bob must be considered as being in a superposed state entangled with the particle and the measurement apparatus. Bob's status is exactly similar to the status of a measuring apparatus. It is important here to notice the difference with what is often assumed in the standard interpretation of the measurement process. For strict realists, the simple fact that the particle interacts with the (macroscopic) measurement apparatus is enough to collapse the wave-function and to record a definite result independently of any observer. This is exactly this assumption that gives rise to the infamous measurement problem. Inside ConSol this is not the case since after the interaction the whole system "particle plus apparatus" is an entangled state and no definite result is obtained. From Alice's point of view, this is even extended to the system "particle plus apparatus plus Bob". This big entangled wave function is a superposition of each possibilities: [result "+" of the spin, apparatus with impact on the upper part of the screen and Bob saying that he saw the result "+"] and [result "-" of the spin, apparatus with impact on the lower part of the screen and Bob saying that he saw the result "-"]. So for Alice no result has been obtained till she communicates with Bob (doing a measurement on him). Then she hangs-on to one of the possible branches where Bob says that he saw one of the possible results.

---

[4] Actually there is a co-emergence of each observer and her empirical world from a more abstract third level. But this is not the place here to develop this point and it has no impact on what we want to say in this paper. A more complete philosophical description will be given in a next paper.
[5] In some way, it is reminiscent of the consistent histories approach of Griffiths[17] and Omnes [18].

As in the Everett's interpretation, there is no physical reduction and the "external" world (i.e. the empirical world) remains totally entangled. But unlike what happens in Everett's interpretation, no observer splits. Even though the (empirical) world remains in an entangled state, each observer perceives one unique result among the many possible ones. This result constitutes her phenomenal world. There is no need that all the observers perceive the same things. That means that each observer perceives her own reality (her own phenomenal world) that is not necessarily the same than the reality of the others. However, as I prove in [7], the formalism guarantees that there cannot be any conflict when an observer asks another observer which result has been gotten about a same measurement. That is why this is called Convivial Solipsism.

What is important to emphasize is that in this framework there is no common reality shared by all the observers. The reality is made for each observer by the totality of the results she got (her phenomenal world) and there is no reason why two observers should have exactly the same, even though there is no possibility for an observer to witness the difference between her reality and the reality of another observer because no observer has a direct access to the perceptions of another observer. The only possibility is to ask another observer what she saw but that is similar to measuring the other (who is in a superposed – entangled - state) and that will provide no information about the real perception she had.

Another important point is that in each observer's reality (her phenomenal world), everything that has not been measured and is not directly related to a branch which the observer is hung-on to is in an indefinite state. This is the key for understanding what we are going to say about the past.

### 2.2 The formal principles

#### 2.2.1. The hanging-on mechanism

Assume that we make a measurement of a given observable P on a system S which is in a state that is a superposition of eigenvectors of P, $\psi_S = \sum c_i |\varphi_i\rangle$ with an apparatus A, initially in the state $\psi_A = |A_0\rangle$. After the interaction between the apparatus and the system, in accordance with the Schrödinger equation, the global system S+A is in the entangled state $\psi_{SA} = \sum c_i |\varphi_i\rangle |A_i\rangle$ where the state $|A_i\rangle$ of the apparatus is correlated to the eigenstate $|\varphi_i\rangle$ of S associated with the eigenvalue $\lambda_i$ of the observable. In standard quantum mechanics with the usual Copenhagen interpretation, the measurement problem comes from the fact that the unitary evolution given by the Schrödinger equation does not allow to get a unique result of the measurement while the reduction postulate says that after the measurement: a) the apparatus is seen in only one of these states, say $|A_{i0}\rangle$, b) the state of the system is projected onto the related eigenstate $|\varphi_{i0}\rangle$ and c) the result of the measurement is the eigenvalue $\lambda_{i0}$ associated with this eigenstate.

If the observer is taken into account, she becomes also entangled with the other systems and according to the Schrödinger equation, the global system after the interaction between the apparatus, the system and the observer is in the entangled state: $\psi_{SAO} = \sum c_i |\varphi_i\rangle |A_i\rangle |O_i\rangle$ ($O_i$ being the state of the observer having seen the result i) and the same difficulty arises. In the absence of any clarification of what a measurement is and when it happens, the problem remains unsolved.

It is then necessary to define precisely what a measurement is. This is what the hanging-on mechanism does:

*"Hanging-on mechanism: A measurement is precisely the awareness of a result by an observer whose perception selects at random (according to the Born rule) one branch of the entangled state vector of the system written in the preferred basis and hangs-on to it. Once the perception is hung-on to one branch, it will hang-on only to branches that are daughters of this branch for all the following observations."*

Consider for example a spin 1/2 particle in a superposed state along Oz.
$$|\Psi_S\rangle = \alpha |+\rangle_z + \beta |-\rangle_z \qquad (1)$$
After the interaction with the apparatus and if we include the state of the observer we get:

$$|\Psi_{SAO}\rangle = \alpha |+\rangle_z |\uparrow\rangle |☺\rangle + \beta |-\rangle_z |\downarrow\rangle |☹\rangle \qquad (2)$$

Where $|\uparrow\rangle$ and $|\downarrow\rangle$ are the states of the apparatus with impact "up" and "down" and $|☺\rangle$ and $|☹\rangle$ are the physical states of the observer's brain correlated respectively to having seen these impacts. Beware to make a difference between the physical brain of the observer and her consciousness. $|☺\rangle$ and $|☹\rangle$ are physical states of the observer's brain. Now, the hanging-up mechanism says that the consciousness of the observer chooses one branch at random. Let's denote by $\widetilde{☺}$ the fact to be aware of having seen "+" (resp. by $\widetilde{☹}$ of having seen "-"). After the hanging-up mechanism, either $\widetilde{☺}$ or $\widetilde{☹}$ will be the case.

We must be very clear not to confuse $|☺\rangle$ with $\widetilde{☺}$ and $|☹\rangle$ with $\widetilde{☹}$. $|☺\rangle$ and $|☹\rangle$ are kets describing the physical states of the observer's brain that enter into the entangled universal sate vector. $\widetilde{☺}$ and $\widetilde{☹}$ are not state vectors and cannot enter into any linear combination with state vectors. That is the reason why they are not written as kets. They are just representing states of awareness.

The entangled global wave function is a description of the observer's empirical world. It is not changed after a measurement. So, it remains as written in equation (2). But inside the phenomenal world which is built through the measurements made on the empirical world, everything happens for the observer as if the wave function was reduced either to $\left|+\right\rangle_z \left|\uparrow\right\rangle \left|☺\right\rangle$ if her state of awareness is $\widetilde{☺}$ or to $\left|-\right\rangle_z \left|\downarrow\right\rangle \left|☹\right\rangle$ if her state of awareness is $\widetilde{☹}$. And this remains true for all subsequent measurements.

I refer the reader to [7] for details about the very meaning of the hanging-on mechanism and the way to show that it is consistent with any predictions of various measurements.

### 2.2.2. The relativity of states

The second assumption is:

"*Relativity of states: Any state vector (and any observable) is relative to a given observer and cannot be considered as absolute.*"

Hence, the global entangled wave function is relative to each observer and so is the empirical world. There is no absolute state vector. This is similar to what is assumed in QBism or in the relational interpretation. Of course, this prevents keeping the traditional image of an external reality which is the same for all the observers. Each observer gets her own state vector for all she is able to observe and that constitutes her empirical world. This state vector evolves deterministically through the Schrödinger equation and remains always a superposition of states of entangled systems. The empirical world is for each observer everything that could potentially give rise for her to an observation[6]. The global (but relative), always entangled, wave function is the description of this empirical reality. The empirical reality evolves only through the Schrödinger equation. After each measurement, the perception of the observer is hung-on to one branch and confined to this branch and its daughters when a new measurement is done. The totality of all the branches the observer is hung-on is her phenomenal world. So the phenomenal world is built from the empirical world through the measurements the observers does and is of course relative too to the observer.

### 2.2.3. The no-conflict theorem

Any transfer of information from an observer to another – for example, any answer made by Bob to a question asked by Alice – unavoidably proceeds through physical means. Therefore it necessary takes the form of a measurement made by A on B.

Hence for every observer, anything outside of her own private perceptions (including any other observer) has to be treated as a physical system that can be entangled with other systems. Suppose that Alice knows that Bob has performed a measurement of the spin along Oz of a spin half particle. Alice has strictly no access to Bob's consciousness. From Alice's point of view, Bob (who is nothing else for her than a physical system similar to an apparatus) is entangled with the particle. So the wave function describing Bob and the particle (after Bob made his measurement) is, for Alice, an entangled wave function with both possible results. Now Alice can perform the same measurement on the particle and Alice will be hung-on as well to one of the two branches and she will see one value. Notice that this branch includes the state of Bob that is linked to the very same value. So when Alice, hung-on to that branch, speaks with Bob to know what Bob saw, she performs a measurement on Bob and, in accordance with the hanging-on mechanism, she cannot hear Bob saying anything else than the value that she has got herself. Alice will never hear Bob saying that he saw "+" when she saw "-". No conflict is possible and the intersubjectivity is preserved. This is why this position is called Convivial Solipsism: each observer lives in her own world but there is room for other observers (contrary to what happens in standard solipsism) and these observers cannot conflict with one another when they communicate.

---

[6] Of course the global entangled wave function describing one observer's empirical world contains everything that could be the subject of a measurement for this observer and it is extremely complex. It is an abstract tool that the observer knows only very partially.

One very important point, I explain in more details in [7], is that inside ConSol each sentence has a meaning only relatively to one unique observer and there is no meta observer above all the observers (sometimes called God's point of view). Therefore the use of sentences mentioning simultaneously the perceptions of two observers, which would express the point of view of a meta observer, is forbidden. This is the reason why the apparent paradox caused by the question: "how is it possible that Alice hears Bob saying that he saw "+" if, in reality, he saw "-"?" cannot be raised because this question mentions simultaneously what Alice and Bob saw and it has no meaning in ConSol.

### 2.3 ConSol, EPR and the measurement of entangled systems

The measurement problem is already cumbersome when only one particle is considered but it becomes even more striking when considering two entangled particles. So let me emphasize the problem of interpretation raised by the measurement of two entangled particles.

The standard experiment for studying this kind of situation is the EPR argument [19]. Two particles A and B in a singlet state of spin are measured by two spatially separated experimenters Alice and Bob:

$$|\psi\rangle = \tfrac{1}{\sqrt{2}}\left[ |+\rangle^A |-\rangle^B - |-\rangle^A |+\rangle^B \right] \qquad (3)$$

In a realist framework where the state is considered as representing the real physical state of the system and where the measurement is assumed to cause the collapse of a superposed state onto one of its components, it seems that the first measurement of one particle (say A by Alice) causes both the collapse of the A state on one of the states $|+\rangle$ or $|-\rangle$ and instantaneously the collapse of the B state on the corresponding opposite state $|-\rangle$ or $|+\rangle$. Then it seems that the measurement on A has an immediate effect on the state of B whatever the distance between A and B be. Avoiding this effect would mean that the states are already determined from the beginning. However this is forbidden by the Bell's inequalities [20] whose violation has been confirmed by many experiments. This is the non-locality of quantum mechanics which is now widely accepted because while apparently violating the theory of special relativity, it is not possible to use it to communicate at a speed faster than the speed of the light because the results of the measurements are random.

Now, the story of this experiment in a realist framework with a measurement causing a physical collapse of the state raises many problems which are left aside too often. The assumptions of the realist stance are first that the state vector corresponds to the physical state of the system (that means that any change in the state vector represents a change in the physical state), second that a measurement projects a superposed state vector onto one of its components (that means that a measurement causes a physical change to the state of the system). So even if it is not possible to use this process to communicate, the measurement on A causes instantaneously a physical change of the state of B. But there is something strange: when the measurements on A and on B are space-like separated, it is impossible to decide which one happened the first. So it is no more possible to say that one measurement causes the result of the other since none can be considered to be the first one. Moreover, even if it is not possible to use entangled particles to communicate at will a message faster than light, such a pair of measurements nevertheless brings a kind of information faster than light. Indeed, if A finds a result "+", she will instantaneously know that B has found the result "-". Of course B cannot use this process to tell instantaneously something particular he has in mind to A, nevertheless the information that he got "-" has been transmitted instantaneously. It is as if a random number generator that is used at one place was able to cause the apparition of exactly the same random number instantaneously at another location. That would be considered as very surprising. This is for example a way to synchronize actions instantaneously between distant points even if these actions have not been decided before. Assume that Alice and Bob are measuring the spin of their particle and that they have agreed on the fact that if the spin is "+" for Alice she will drink a cup of tea and if it is "-" she will drink a glass of wine (and the same for Bob). So, when they do their measurements, they instantaneously know what the other one is doing. For example, the information that Alice is drinking a cup of tea has been communicated instantaneously to Bob. Of course Alice cannot decide by herself what she is going to do and then send the information to Bob. However, Bob is able to know instantaneously what she is doing even though what she does has been decided just a very short moment before and that Bob is at a distance of 1 billion light years. And this can be improved to communicate much more elaborate pieces of information depending on the protocol Alice and Bob have agreed on and the number of measurements they use. They can link complicated information on things done at each location depending on strings of "+" and "-" gotten in the measurements. Then it is not totally true to say that this process cannot be used to communicate any information even if it is true that this rests on a conventional agreement made before and that what happens at each location is not deliberate but decided at random by the results of the measurements.

As I say in [1]:

*"If a random number generator that is used at one place was able to cause the apparition of exactly the same random number instantaneously at another location, it would not be possible to use it to communicate but that would nevertheless be considered as violating Relativity. So for many physicists a way to avoid the problem is often to say that in the EPR case, we notice a mere correlation between*

*two results and to claim that correlation is not causality. But this is not an acceptable reason since in classical statistics the precise reason for the difference between correlation and causality is the fact that a common cause can be invoked which is forbidden here by the Bell's inequalities".*

Hence one sees that there is a real problem in interpreting the collapse as a real change in the physical state of the system even though a large number of physicists decides to forget it.

Now let's come back to the EPR experiment and try to understand better what happens. In particular, using ConSol, we can explain first, why the order in which the measurements are made is irrelevant and second, that there is no need to invoke any non-locality[7]. Alice and Bob measure each one the spin along the same axis of one of the two particles of an entangled pair in a singlet state:

$$|\psi\rangle = \tfrac{1}{\sqrt{2}}\left[\,|+\rangle^A\,|-\rangle^B - |-\rangle^A\,|+\rangle^B\,\right] \qquad (4)$$

We know that inside the framework of ConSol we cannot speak of a measurement without mentioning who is the observer. Let's adopt Alice's point of view. The fact that Bob does a measurement on his particle means that he becomes entangled with it and from Alice's point of view the global state becomes:

$$|\psi\rangle = \tfrac{1}{\sqrt{2}}\left[\,|+\rangle^A\,|-\rangle^B\,|☺\rangle^B - |-\rangle^A\,|+\rangle^B\,|☹\rangle^B\,\right] \qquad (5)$$

When she does a measurement on her particle at time $T_0$, she can find either "+" or "–" depending on the branch she hangs-on. Nothing physical happens to her particle nor to Bob's particle. They both stay entangled in Alice's empirical world but for Alice a result, let's say "+", has been obtained in her phenomenal world and she is hung-on to the corresponding branch: $|+\rangle^A\,|-\rangle^B\,|☺\rangle^B$. When later, at time $T_1$, she talks to Bob, she will necessarily hear him saying that he got the result "-". In this simple example, the result that Alice gets at $T_0$ determines one unique result for the measurement she will make on Bob. So, it is natural to think that "Bob got the result "-"" was already true for Alice at $T_0$[8]. Indeed, in this simple case where the result is unequivocally determined by Alice's first measure, everything happens for Alice as if everything becomes determined at $T_0$ immediately after her first measurement. This is the reason why non-locality is invoked. Indeed, assume that Bob's and Alice's measurements are separated by a space-like interval, then it seems that Alice's measurement resulted in changing Bob's particle state faster than what is allowed by special relativity. But this interpretation is wrong since Alice's measurement has no physical impact on Bob's particle state. What happens is simply that she is hung-on to one branch that includes the right correlation. This is also the key to solve the famous puzzle raised by the fact that if we think that the first measurement is the cause of the result of the second one, then it is very hard to understand how this can be the case when the two measurements are space-like separated since none can be said to be the first. Is Alice's measurement and the result she finds ("+") the cause of Bob's measurement result ("-") or is Bob's measurement and the result he finds ("-") the cause of Alice's measurement result ("+")? And above all, since each measurement can be considered to be the first and that each measurement gives a result at random, how is it possible that the two results remain correlated? This is clearly very weird! Inside Consol no such strange thing happens. From Alice's point of view, once hung-on to one branch, there is no surprise that she hears Bob saying that he found the result correlated to her's and independently, from Bob's point of view, that is the same thing. There is no more any ambiguity about which measurement is the first since the observer from which the point of view is considered does herself both measurements. The first one results in a hanging-on to one branch. The second one simply notices the unique result correlated in this branch for the other particle. Let's keep in mind that inside Consol, there is no need that Bob "really" saw what Alice hears him to say he saw and that Alice "really" saw what Bob hears her to say she saw. That is even meaningless. There is no possibility to compare in an absolute way what they saw. So the mystery of space-like separated measurements and of the question to know which one is the cause of the other disappears. The most important point is to keep in mind that nothing physical happens during these measurements but only a hanging-on of each observer to one of the possible branches and not necessarily the same for both. So there is no longer any non-locality understood as something causing a physical change at a space-like separated location. This will be important for understanding the way we consider past events in part C.

---

[7] For a detailed discussion of this point see [2].
[8] We will analyze later more precisely that kind of events especially in the more interesting case where Alice's first measurement does not determine unequivocally the result she gets for the second measurement (when she talks to Bob to know which result he got). In this case, it will no more be possible to make the confusion letting believe that "Bob got such and such result" was already true for Alice at $T_0$.

### 3. Part B: The delayed choice experiments

It is well known that Wheeler proposed a set of so called "delayed choice" experiments to try to analyze when exactly a photon "decides" which behavior it is going to adopt in a an experiment where the choice of the device that is used is decided at the very last moment before the measurement [21, 22]. As famously emphasized by Feynman [23], in the classical double-slit experiment the photon seems to be able to choose which behavior (wave or particle) it is going to adopt depending on the whole experimental setup that is in place. This is already strange enough. But in the case of a delayed choice experiment, it is even stranger since it seems that changing the experimental device at the very last moment has an effect on what the photon decided well before. This is the reason why backward in time communication is sometimes invoked to explain the results [4].

In [1] I separate two different situations: the first one concerns experiments involving only one particle and the second one, experiments involving two entangled particles.

#### 3.1 Experiments involving only one particle

Let's briefly recall the analysis of the first situation. All the experiments of this kind (including those using a Mach-Zehnder interferometer) are similar to a double slit experiment where it seems that the choice between either detecting the slit the particle went through or not detecting it can be done at the very last moment (see figure 1).

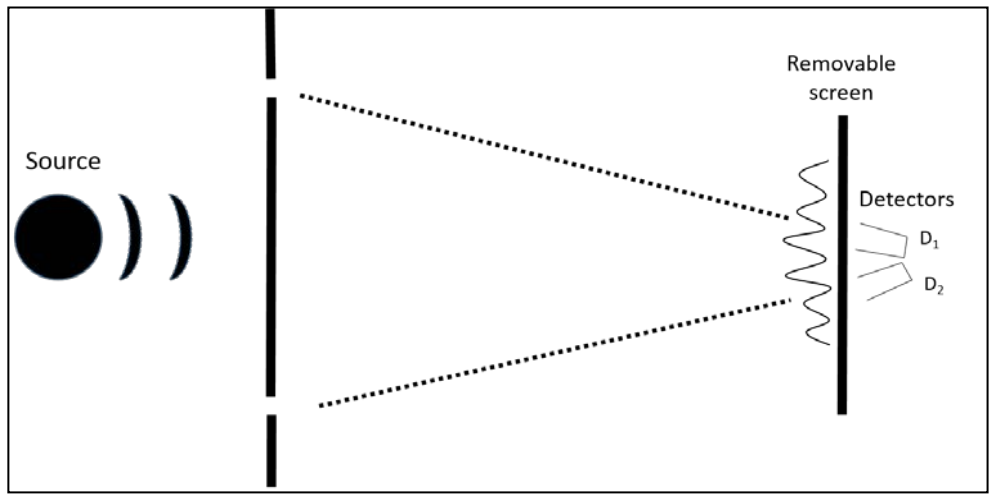

**Fig. 1: Delayed choice double slit experiment**

Particles are sent one by one by a source. Either a screen (removable) receives the particle or two detectors (D1, D2), each one pointing precisely at one slit, receive it. When the screen is present an interference pattern is seen (interpreted as a wave behavior). When the screen is removed, each particle triggers only one detector allowing (as alleged) to know which slit the particle went through (interpreted as a particle behavior). It is then tempting to say that when the screen is present (interference), each particle decides to pass through both slits while when it is removed (no interference), each particle chooses only one slit to pass through. So, it seems that the particle is "informed" of the presence or of the absence of the screen (no matter how far away it is) at the very moment when it passes the slits. A delayed choice situation corresponds to the case where choosing to put or remove the screen is done well after the particle went through the slits. While in the former case, it is still possible to think that the "information" about the design of the device is brought to the particle by a standard (though totally unknown) process travelling at a speed lower than the speed of the light, this is no more possible in the latter case where the design is decided at the very last moment and after the "choice" that the particle has to make. Here the conclusion seems to be that putting or removing the screen has an influence in the past on the decision of the particle to go through only one or both slits. Such an experiment (using a Mach-Zehnder interferometer) has been done recently and the authors conclude [3]:

*"Our realization of Wheeler's delayed choice GedankenExperiment demonstrates beyond any doubt that the behavior of the photon in the interferometer depends on the choice of the observable which is measured, even when that choice is made at a position and at a time such that it is separated from the entrance of the photon in the interferometer by a space-like interval. In Wheeler's words, since no signal*

*travelling at a velocity less that of light can connect these two events, "we have a strange inversion of the normal order of time". [...] Once more, we find that Nature behaves in agreement with the predictions of Quantum Mechanics even in surprising situations where a tension with Relativity seems to appear."*

This interpretation is wrong. I briefly sum up here the analysis done in [1]. The wave function of the particle is always the sum of two spherical waves before reaching either the screen or the detectors[9].

$$\psi(\vec{r}) = \frac{1}{|\vec{r}-\vec{r_1}|} e^{ik|\vec{r}-\vec{r_1}|} + \frac{1}{|\vec{r}-\vec{r_2}|} e^{ik|\vec{r}-\vec{r_2}|} \qquad (6)$$

where k is the wavenumber of the particle and $\vec{r_1}$ and $\vec{r_2}$ are the locations of the two slits.

The mistake causing the astonishment is simply to think that when a detector clicks, it indicates which slit the particle went through and to deduce from that that it had a particle behavior. That is wrong. Indeed, it is possible to approximate (1) as a sum of two planar waves:

$$\psi(\vec{r}) \sim \frac{1}{|\vec{r}-\vec{r_1}|} e^{i\vec{k_1}(\vec{r}-\vec{r_1})} + \frac{1}{|\vec{r}-\vec{r_2}|} e^{i\vec{k_2}(\vec{r}-\vec{r_2})} \qquad (7)$$

where $\vec{k_1}$ and $\vec{k_2}$ are wavevectors of size *k* directed along $\vec{r_T} - \vec{r_1}$ and $\vec{r_T} - \vec{r_2}$ respectively and $\vec{r_T}$ is the location of the detectors. So, if we remember that the detectors measure the momentum of the incoming particle, with a very narrow detection range, a click of detector $D_1$ (resp. $D_2$) is a measurement of the momentum around $\vec{k_1}$ (resp. $\vec{k_2}$) which, according to the Born rule, happens with a probability ½ when the wave function is (7). It is important to understand clearly what the mistake is: One assumes that because the detector $D_1$ clicks - hence that the measurement of the momentum gives the result $\vec{k_1}$ - the particle went through the slit 1 because $\vec{k_1}$ points towards slit 1. But that is implicitly assuming that the momentum was already $\vec{k_1}$ before the measurement, which is not true. The value of the momentum is determined only after the measurement. So from the fact that after the measurement the momentum is $\vec{k_1}$, it is not legitimate to infer that the particle went through the slit 1. The state of the particle before the measurement is exactly the same - described by equation (1) or its approximation equation (2) - whatever the design be (screen or detectors) and there is absolutely no change of the state of the particle before the measurement. It is only when a measurement is made that the state changes and is projected on a definite position if the measurement is made by the screen or a definite momentum if the measurement is made by the detectors. Hence, there is no (backward in time or not) effect of the design of the device on the wave function before the particle reaches the screen or the detectors.

---

[9] I give a detailed computation in [1].

Notice that this simple explanation can perfectly be given in a standard realist framework where the state of the particle can be considered as describing a real physical state and where the measurement is interpreted as in standard quantum mechanics through the reduction postulate and the Born rule.

### 3.2 Experiments involving two entangled particles

#### 3.2.1 The delayed choice quantum eraser (point of view 1)

Let's now consider the delayed choice quantum eraser experiment. Many experiments have been made[10] and we will focus on one of them [25]. In this experiment (see figure 2), light is sent through a double slit and goes through a BBO crystal for a parametric down conversion splitting each photon in two entangled photons. $D_0$ is a moveable detector playing a role similar to a screen to detect an interference pattern for the photons travelling upward (signal photons). The photons travelling downward (idler photons) are received by a prism and a set of beam splitters (BS) and mirrors (M). It is assumed that if a photon is detected by $D_1$ or $D_2$, it may have travelled via either of the two routes. If it is detected by $D_3$ or $D_4$, it can only have come through one route. The experimental set up is such that the signal photon is detected by $D_0$ before the detection by one of the other detectors of its idler twin photon.

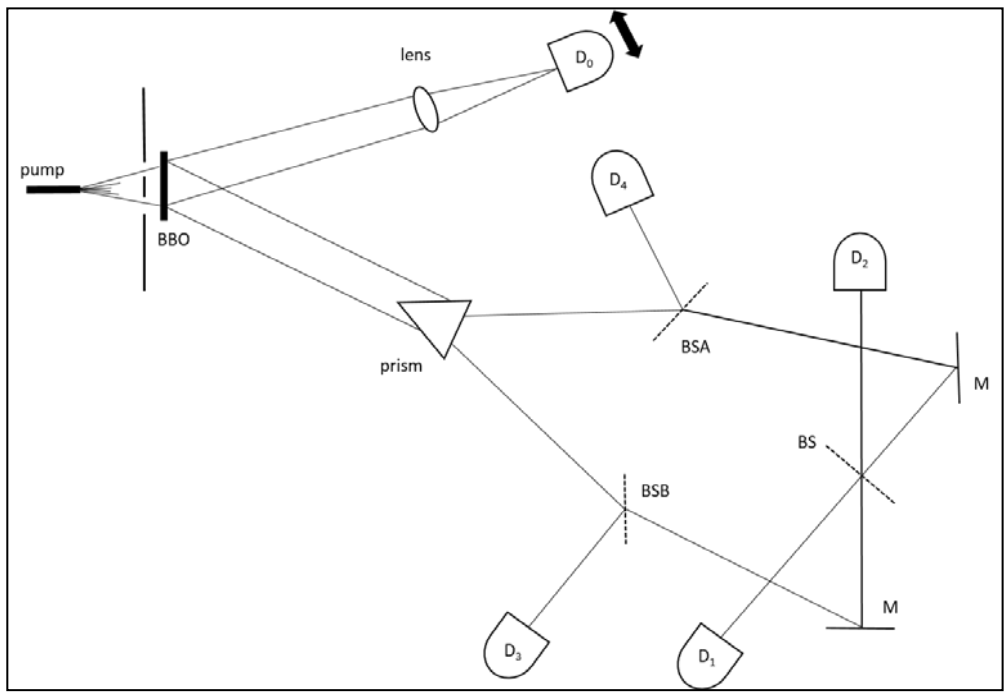

**Fig. 2: Delayed choice quantum eraser**

When the results of the detectors are compared through the joint detection rate (the time delay between a triggering of $D_0$ and a triggering of $D_{i\ (i=1\ to\ 4)}$ is constant), one sees that the signal photons associated with idler photons detected by $D_1$ or $D_2$ show an interference pattern while the others show no interference.

---

[10] See [24] for a recent review.

Indeed, when the beam splitter BS is present, it is not possible to know which slit the photons detected by $D_1$ and $D_2$ came from. If it is removed, a click of $D_1$ (resp. a click of $D_2$) is associated with a photon having gone through the upper slit (resp. the lower slit) and the interference pattern is lost.

The strange thing is that it is possible to put or to remove the beam splitter BS after that the signal photon has been registered by $D_0$. Then, remembering the discussion about the double slit delayed choice experiment, if we adopt the (usual but wrong) idea that the photon went through both slits when it shows interferences and went through only one when it does not, it seems here that deciding to put or to remove BS has an effect on the behavior of the signal photon well before the decision is made.

Nevertheless, as I emphasize in [1]:

*"One important point to notice is that it is possible to recognize an interference pattern only through the extraction, from all the detections registered by $D_0$, of those corresponding either to detections by $D_1$ or to detections by $D_2$. That has the consequence that it is impossible to use this device to transmit information from the future to the past because recognizing an interference pattern is possible only when one have the knowledge of which detection by $D_0$ corresponds to which detector $D_i$. In particular, the tempting idea consisting in putting or removing the beam splitter BS in the far future to produce interference (BS present) or not (BS absent) in the present, is not working because there is no way to distinguish an interference pattern from no interference just by looking at $D_0$ since the interference patterns correlated to $D_1$ and to $D_2$ have a $\pi$ phase shift producing jointly exactly the same image as no interference at all. Hence it is impossible to separate them without knowing which impact corresponds to which detector and this necessitates information that can be known only after deciding to put or not the beamsplitter."*

I show in [1] that the global wave function can be written as:

$$|\psi\rangle = \frac{1}{2\sqrt{2}} |1\rangle^{\searrow} [\,|U\rangle + |L\rangle\,]^{\nearrow} + \frac{1}{2\sqrt{2}} |2\rangle^{\searrow} [\,|U\rangle + |L\rangle\,]^{\nearrow} + \frac{1}{2} |3\rangle^{\searrow} |L\rangle^{\nearrow} + \frac{1}{2} |4\rangle^{\searrow} |U\rangle^{\nearrow} \qquad (8)$$

where the arrows indicates photons going up (signal photons) or down (idler photons), $|U\rangle$ and $|L\rangle$ indicates photons went through the upper hole or the lower hole and $|i\rangle$ is an abbreviation for the combined states of "photon downward arriving at detector $i$ and detector $i$ clicking".

This equation is similar to a EPR situation. It indicates that $D_1$ (or a $D_2$) clicks are correlated to the $\frac{1}{\sqrt{2}}[\,|U\rangle + |L\rangle\,]^{\nearrow}$ part of the wave function. Hence these clicks will be correlated to signal photons giving rise to an interference pattern. $D_3$ (resp $D_4$) clicks are correlated to the $|L\rangle^{\nearrow}$ (resp. $|U\rangle^{\nearrow}$) part of the wave function. Hence these clicks will be correlated to signal photons giving rise to no interference pattern.

Now if the beam splitter is removed the global state becomes:

$$|\psi\rangle = \tfrac{1}{2} |1\rangle^{\searrow} |U\rangle^{\nearrow} + \tfrac{1}{2} |2\rangle^{\searrow} |L\rangle^{\nearrow} + \tfrac{1}{2} |3\rangle^{\searrow} |L\rangle^{\nearrow} + \tfrac{1}{2} |4\rangle^{\searrow} |U\rangle^{\nearrow} \qquad (9)$$

And we see that there is no more any interference for the signal photon.

A wrong (but tempting) reasoning could be to say that since it is possible to decide to let or to remove the beam splitter after that the signal photon has been detected, a signal photon that does not interfere and that is correlated to an idle photon detected by $D_1$ or $D_2$ in the case where the beam splitter has been removed behaves as if it knew in advance that the beam splitter was going to be removed. But notice that it is impossible to know if one isolated photon is part of an interference pattern or not. The only way to see an interference pattern is to analyze a whole set of impacts corresponding to many photons. In this experiment, it is impossible to detect an interference pattern without extracting the points that are related to one detector. This is because, as I said above, there is no way to distinguish an interference pattern from no interference just by looking at $D_0$ since the interference patterns correlated to $D_1$ and to $D_2$ have a $\pi$ phase shift producing jointly exactly the same image as no interference at all and it is impossible to separate them without knowing which impact corresponds to which detector.

The interpretation of the experiment is simple in the framework of Convivial Solipsism if we adopt, as we did above, the point of view of an observer Alice who uses first the detectors $D_{i(1,\ldots 4)}$ and then asks Bob which results he got from the signal photons at $D_0$. According to the hanging-on mechanism, when Alice learns about the results concerning the signal photons this must not be interpreted as Alice learning a certain state of affairs already fixed but as Alice making a measurement on signal photons (or on Bob who is entangled with them). Indeed for Alice, Bob's measurements do not produce a definite result but only an entanglement between Bob and the photons he measures. Then for each signal photon, she can get a result coming only from the branch she is hung-on to after her measurement on the corresponding idler photon. Hence, in the case where the beam splitter is present, she must use the wave function (8) and for all the signal photons $j$ corresponding to idler photons for which she got the result 1, she is hung-on to $\left\{ \tfrac{1}{\sqrt{2}} |1\rangle_j^A \left[ |U\rangle + |L\rangle \right]_j^B \right\}$ and so she gets interference when she measures them (idem for result 2). For all the signal photons k corresponding to idler photons for which she got the result 4, she is hung-on to $\left\{ |4\rangle_k^A |U\rangle_k^B \right\}$ and so she gets no interference (idem for 3). Now if the beam splitter is removed, she must use the wave function (9) and for all the signal photons k corresponding to idler photons for which she got the result 1 or 4, she is hung-on to $\left\{ |1\rangle_k^A |U\rangle_k^B \right\}$ or $\left\{ |4\rangle_k^A |U\rangle_k^B \right\}$ while for all the signal photons $j$ corresponding to idler photons for which she got the result 2 or 3, she is hung-on to $\left\{ |2\rangle_j^A |L\rangle_j^B \right\}$ or $\left\{ |3\rangle_j^A |L\rangle_j^B \right\}$. Hence there will be no interference.

This explanation that I gave in [1] is mainly focused on the perspective of an observer doing firstly measurements of idler photons through the detectors $D_{i\,(i=1,...4)}$ and only in a second time measuring the position of the signal photons. The perspective of an observer doing firstly measurements of signal photons and second, noticing for each one which detector its twin idle has triggered was not considered. It is true that I formally reduced the quantum eraser to a EPR situation and that it is well known that the order of measurements in this type of experiments is irrelevant. Hence the two perspectives should be equivalent. But in a EPR situation there is a one to one correspondence between the results obtained for the two particles. This is not the case with the quantum eraser and the situation is more complex if we adopt the point of view of an observer using first $D_0$. So we present now a new explanation of what happens when the position of the signal photons are observed well before the decision to put or to remove the beam splitter is made.

### 3.2.2 *The delayed choice quantum eraser (point of view 2)*

Let's switch to Bob's position, the observer that measures the signal photons. Things could be presented like that: Assume Bob does his measurements during a certain period and that, due to the design of the experimental device, the idler photons arrive in the area of the detectors $D_1$ to $D_4$ only a long time after the measurement of the last signal photon by Bob. Assume as well that the decision to put or to remove the beam splitter is made at random just a minute before the first idler photon enters the area of detection. Let's remember that if the beam splitter is present then signal photons associated with idler photons detected by $D_1$ and $D_2$ will show an interference pattern while if there is no beam splitter, there will be no interference. So, how can a signal photon detected by $D_0$ at $T_0$ and associated with an idler photon detected by $D_1$ at $T_1 \gg T_0$ know if it must interfere or not at $T_0$ since it is detected well before the decision to put the beam splitter is made (a very short time before $T_1$)?

From this side, the question seems even more puzzling than from the point of view of the observer using $D_1$ to $D_4$. The astonishment comes from the fact that we reason as if it was possible to know that a single photon has interfered with itself just by looking at its recorded position. If this was possible, then that would mean that it is possible to send a message from the future to the past just by deciding in the future to put or to remove the beam splitter and the result would be that the corresponding signal photon would interfere or not, letting the observer in the past know what the decision in the future will be. But that is not possible as we have already explained.

It is impossible to know if one isolated photon is part of an interference pattern or not. The only way to see an interference pattern is to analyze a whole set of impacts corresponding to many photons. And even worse, in this experiment it is impossible to detect an interference pattern without extracting the points that are related to one detector ($D_1$ or $D_2$) since the interference patterns correlated to $D_1$ and to $D_2$ have a $\pi$ phase shift producing jointly exactly the same image than no interference at all. So there is no way to distinguish an interference pattern from no interference just by looking at $D_0$.

That is the key to understand what happens. Bob got all his results from $D_0$ and has no way to know which photons are part of an interference pattern and which are not. All what he sees is a fuzzy picture of photons with random positions. It is only when he will be able to associate each signal photon with its corresponding idler photon that he will be able to select among all the positions he got from $D_0$ those which are related to idler photon detected by $D_1$ (or by $D_2$). This selection will then allow him to see if there is an interference pattern or not. Is the mystery solved? Yes as we are going to show.

The important point is to notice that to each signal photon detected at a certain position, it is necessary to associate the detector triggered by its idler brother. According to what we saw about the measurement of entangled particles, the measurement of the first particle (the signal photon) results in a hanging-on of the observer to the branch of the global wave function which is associated to the result obtained. In the classical EPR experiment the branch corresponding to one result for the first particle allows only one result for the second particle ("up" for the first particle implies "down" for the other and vice versa). But in more complex cases such as the quantum eraser, one result for the first particle can left undefined the result for the second particle if the wave function is such that it contains the tensorial product of an eigenstate of the observable that is measured on the first particle and of a superposition of eigenstates of the observable that is measured on the second particle. In this case, once the first measurement has been made, the observer is hung-on to the corresponding branch and getting the result of the measurement of the second particle is equivalent to measure the second particle in the superposed state of this branch.

A very simple example to illustrate this point is given by two entangled particles A and B which have the following global state written in the basis which is the tensorial product of the basis of eigenvectors of the observable measured on A and the basis of eigenvectors of the observable measured on B:

$$|\psi\rangle = \frac{1}{\sqrt{3}}\left[\,|a\rangle + |b\rangle\right]^B \left[\,|X\rangle\right]^A + \frac{1}{\sqrt{3}}\,|a\rangle^B\,|Y\rangle^A \qquad (10)$$

If Alice measures A first she has a probability 2/3 to find X and a probability 1/3 to find Y. If she finds Y, she will be hung-on to the branch $|a\rangle^B\,|Y\rangle^A$ and when she will measure B she will necessarily find a. This is what happens in the EPR experiment. Finding a result for one particle fixes the result for the other. Now if she finds X, she will be hung-on to the branch $\frac{1}{\sqrt{2}}\left[\,|a\rangle + |b\rangle\right]^B \left[\,|X\rangle\right]^A$ and when she will measure B, she will have an equal probability ½ to find a or b. Knowing the result of the measurement on A does not determine the result of the measurement on B which remains undefined until a real measurement is made.

Another important point to notice is the fact that the global entangled wave function one must use to determine the possible results of the second measurement is the wave function that corresponds to the final experimental device. So if the beam splitter is present the wave function to be used is:

$$|\psi\rangle = \frac{1}{2\sqrt{2}}\big[|1\rangle + |2\rangle\big]^{\searrow}\big[|U\rangle + |L\rangle\big]^{\nearrow} + \frac{1}{2}|3\rangle^{\searrow}|L\rangle^{\nearrow} + \frac{1}{2}|4\rangle^{\searrow}|U\rangle^{\nearrow} \tag{11}$$

While if it is removed, the wave function to be used is:

$$|\psi\rangle = \frac{1}{2}|1\rangle^{\searrow}|U\rangle^{\nearrow} + \frac{1}{2}|2\rangle^{\searrow}|L\rangle^{\nearrow} + \frac{1}{2}|3\rangle^{\searrow}|L\rangle^{\nearrow} + \frac{1}{2}|4\rangle^{\searrow}|U\rangle^{\nearrow} \tag{12}$$

That is a fundamental point because it is the reason why, from exactly the same list of detections by $D_0$, an interference pattern will be associated to $D_1$ (or $D_2$) or not. These two wave functions tell us how to link a photon detected at a certain position by $D_0$ to the detectors $D_{i\,(i=1...4)}$. Of course the signal photons which have already been recorded in $D_0$ do not care what happens in the future to their idle twin. The list of detections by $D_0$ can be completed well before the first detection by one of the $D_{i\,(i=1...4)}$ is made. And what will be decided about putting or not the beam splitter has absolutely no influence at all on what the signal photons do. But putting or removing the beam splitter is the crucial point that decides which one of the two entangled wave functions is the relevant one for the measurement on the idler photon which is done only after. In other words, putting or removing the beam splitter tells which branch Bob will be hung-on for the measurement of the twin idler photon after having measured the position of a signal photon. For each signal photon the result that will be found for its idler twin depends on the branch the observer is hung-on and this branch is not the same depending on which wave function is the good one. For each signal photon, the probability that it is associated to an idle photon detected by one of the $D_{i\,(i=1...4)}$ will not be the same for (11) and for (12). Using the wave function (12) corresponding to no beam splitter will give for each photon of the set of impacts described by $|U\rangle$ (resp. $|V\rangle$) an equal probability to be associated with $D_1$ or $D_4$ (resp. $D_2$ or $D_3$). The signal photons will be uniformly distributed between the detectors, hence there will be no interference pattern seen for any of the photons associated with one detector. Using the wave function (11) corresponding to the presence of the beam splitter will in contrast result in a higher probability to link certain signal photons with $D_1$, certain others with $D_2$ and the distribution of the signal photons to the detectors will no more be uniform. For signal photons at positions corresponding to some fringes the probability that their twin be associated to $D_1$ (resp. $D_2$) will be larger than the probability to be associated to $D_2$ (resp. $D_1$). The consequence will be that when considering only signal photons filtered by $D_1$ (resp $D_2$), they will show an interference pattern.

So the interference patterns are created by the post selection process of the second measurements and are in no way present before! It is just that using the distribution function which links a signal photon to a detector $D_i$ can be done in two different ways through two different wave functions which give for the first one a pattern of interference for the photons filtered by $D_1$ (resp. $D_2$) and no pattern for the second one, from exactly the same initial list of signal photons gotten from $D_0$. The interference pattern is created during the last measurements which link signal photons to detectors through the wave function (11) and not before.

### 3.2.3 Summary of the two points of view

We can summarize our conclusions as follows:

From Bob's point of view (Bob measures first the signal photons) the signal photons are recorded well before he measures the idler photons and well before the decision to put or to remove the beam splitter is made. Putting or removing the beam splitter has absolutely no effect on the positions of the signal photons but it determines which branch Bob is hung-on when later he measures idler photons for each recorded position of a signal photon. Of course it is obvious that putting or removing the beam splitter has an impact on the idler photons since that changes their path. Hence that results in determining the correct wave function to use for the probabilities of the measurements of the idler photons and which branch Bob is hung-on[11]. According to these probabilities that links each signal photon to an idler photon detected by one detector, the filtering of the signal photons associated with $D_1$ (or $D_2$) will give or not an interference pattern. Hence there is no effect from the future to the past but only a filtering depending on the presence or the absence of the beam splitter and that decides through the last measurements which signal positions must be considered as belonging to the same set and so, makes appearing an interference pattern or not.

From Alice's point of view (she uses first the detectors $D_{i(1,...4)}$ on idler photons), when she asks Bob which results he got from the signal photons at $D_0$ she can only get results coming from the branch she is hung-on to. Assume that the beam splitter is present. If she got the result 1, she is hung-on to $\frac{1}{\sqrt{2}}|1\rangle[\,|U\rangle + |L\rangle]$. So before she asks Bob, the position of the signal photons at $D_0$ are not defined for her and that, even though the measurement by Bob through $D_0$ is supposed to have been made well before. Does that mean that there is an effect from the future to the past? Not in the usual sense of a modification of the physical state of the signal photons in the past inside a reality shared by all the observers. That was the conclusion of [1]. But we can be more precise now. In Alice's phenomenal reality, the position of the signal photons are undefined even though they are supposed to have been measured by Bob before. For Alice, Bob's measurements are not measurements but only entanglements between Bob, the apparatus and the system. It is only when Alice speaks to Bob that this becomes a real (for her) measurement and that the position becomes defined. This is an example of past events becoming defined well after they are supposed to have happened. We will analyze more precisely that sort of events in the next paragraph. So when Alice asks Bob about the results he got on the photons for which she saw a $D_1$ click, she makes a measurement on photons in a superposed state and she gets interference (idem for $D_2$). One sees now that the explanation given in [1] is easier to understand when completed with the concept of phenomenal reality limited to one observer.

---

[11] To be more precise, if there is no beam splitter when Bob does his measurements on signal photons he will be hung-on to one of the components of the wave function (12). If the decision is made latter to put the beam splitter, this will change the branch Bob is hung-on to. This is normal since adding a beam splitter is a physical change of the experimental device used for the last measurement.

## 4. Part C: Events and past events

### 4.1.1 What is an event?

We have shown through the analysis of the delayed choice quantum eraser how past results are sometimes determined only by a measurement done afterward, though that should not be confused with a backward in time effect. This allows us to generalize this feature which is not limited to the delayed choice experiments.

All what is said in the following about events is to be understood as events happening in one observer's reality, in her own phenomenal world. Events are not shared between observers as if they were belonging to a common universal reality (which does not exist). An event is either a result that the observer got directly from a measurement at some time or is related to a measurement that the observer has done. An event is related to a measurement when it belongs to the same branch than the branch to which the observer hangs-on through her measurement and that this measurement determines unequivocally the corresponding value. That means that the event is related to a property of a system that is entangled with the system that is measured. A good example is the case of EPR correlations in a singlet state. When Alice measures her particle and finds "+" the fact that her particle has a spin "+" is an event directly observed. The fact that the other particle has a spin "-" is also an event even though Alice did not directly measure the spin of this particle, since the value of its spin is determined by the first measurement Alice did. In this simple case, Alice hangs-on to a branch that contains only one possibility for the other entangled system. But that is not necessary and the two systems could be entangled in such a way that the branch Alice hangs-on to after her first measurement contains a superposed state of the second system. This is the case of the wave function (10) if Alice gets X after her first measurement. In this case, after the first measurement there would be no event concerning the second system. Hence there is no absolute (in the sense of observer independent) event. An event is always related to one particular observer and is an event only for this observer. To paraphrase a famous Wheeler's sentence[12]:

*"No event is an event until it is an observed event"*.

The reality of each observer, her phenomenal world, is the sum of all the events for this observer. There is a consequence on the meaning of the word "true". Truth is no longer something that is absolute and valid for everybody but is relative to each observer inside her phenomenal world. For each observer, is true what corresponds to an event inside her phenomenal reality. This is still a correspondentist notion of truth because the truths makers are for each observer as real as any truth makers in the usual conception of truth. The only difference is that while the truth makers are shared in the usual conception, they are private in convivial solipsism. But there is no way to see the difference for anybody. In the usual conception, each observer thinks that the truth makers for him are the same than for the others. But what is the proof of that? It's just a matter of faith. The important point is that for each observer, everything happens in Convivial Solipsism exactly as it would happen if the standard realism was the correct position. There is no way to witness the fact that our phenomenal world is private. Everything happens as if it was shared and that is the reason why the standard realism is so strongly anchored in our mind.

---

[12] The real sentence is: "*No elementary phenomenon is a phenomenon until it is a registered (observed) phenomenon*" [22].

Let's come back again to the measurement process according to ConSol. Assume Alice does a measurement of the Oz component of a spin half particle. The state of the system is:

$$|\Psi_S\rangle = \alpha |+\rangle_z + \beta |-\rangle_z \tag{13}$$

After the interaction with the apparatus the global state is:

$$|\Psi_{SA}\rangle = \alpha |+\rangle_z |\uparrow\rangle + \beta |-\rangle_z |\downarrow\rangle \tag{14}$$

where $|\uparrow\rangle$ (resp. $|\downarrow\rangle$) is the state of the apparatus with a needle up correlated to a spin + (resp. a needle down correlated to a spin –).

It is only when Alice has a look at the apparatus and perceives one of the two possibilities that the measurement (for Alice) is completed. Before that, Alice must consider that the system and the apparatus are in an entangled state and that no definite result obtains. Hence neither the system nor the apparatus are in a state that can be interpreted classically. In particular, the needle of the apparatus in neither up nor down. This is a major difference with the standard interpretation where it is hardly accepted that macroscopic objects can be in a superposed state[13]. When Alice has a look at the apparatus, she can perceive only one of the two possibilities (needle up or needle down). Then she hangs-on to one of the two branches and the measurement is done: for her, the needle of the apparatus is up (resp. down) and the spin of the particle is "+" (resp. "-"). So "the spin of the particle is "+"" is now an event for Alice. After that, she will stay hung-on to this branch for all subsequent measurements. But notice again that the physical state of the system and the apparatus has not changed and remains entangled inside Alice's empirical world. It is only the perception that Alice has of this state (in her phenomenal world) that is now filtered and limited to only one branch.

Let's now analyze again what happens in the situation where Alice asks Bob about a measurement Bob did (let's say once more a measurement of the Oz component of a spin half particle). For Alice, after that the system and the apparatus have interacted and that Bob has given a look at the apparatus but before she talks to Bob, the whole system is in the state:

$$|\Psi_{SAO}\rangle = \alpha |+\rangle_z |\uparrow\rangle |☺\rangle + \beta |-\rangle_z |\downarrow\rangle |☹\rangle \tag{15}$$

---

[13] For the sake of simplicity, we don't mention here the decoherence which is analysed carefully in the context of ConSol in [7].

where $|☺\rangle$ (resp. $|☹\rangle$) stands for Bob's state corresponding to having seen the needle up (resp. the needle down). Let us remind that from Alice's point of view, Bob has the same status than an apparatus. It is only a physical system. The fact that Bob has perceptions is irrelevant for Alice. Hence, before she talks to him (which is exactly equivalent to having a look at the needle of the apparatus in the previous situation), she must consider that the system, the apparatus and Bob are in the entangled state (15). For Alice, the fact that Bob has made a measurement (i.e. has perceived a result) does not result in a definite result for her. Bob's measurement has exactly the same status than the interaction between the system and the apparatus. It is just an interaction between two physical systems resulting in an entanglement between Bob and the apparatus and for Alice no definite result obtains. It is only when Alice talks to Bob (which is the equivalent of measuring Bob) that she will hang-on to one of the two branches and will hear him saying that he has gotten a result (say "+"). At this very moment, the measurement will be completed and Alice will be hung-on to the branch where Bob has seen "+", the needle of the apparatus is up and the spin of the particle is "+". It is only at this moment that this will become an event for Alice.

### 4.2 Past events

Let's now give a more precise description about when an event happens. Assume that Monday at noon, when talking to Bob, Alice hears him saying that he did his measurement Sunday at eleven. What counts as an event for Alice in this story? First, what Alice perceives directly is an event: hearing Monday at noon Bob giving the result he got is an event: "Bob says that he got the result "+" " is an event dated Monday at noon for Alice because she perceives it directly Monday at noon. So it is an event whose time is equal to the moment (i.e. which happens) when Alice becomes aware of it. Let's go further: Monday at noon, she becomes hung-on to the branch where Bob says that he saw "+" Sunday at eleven and that the needle of the apparatus was up and that the spin of the particle was "+". Now, the information contained in this branch (Bob made a measurement Sunday at eleven, the needle of the apparatus was up and the spin of the particle was "+") is composed by events such that the time indicated for the moment they happened is Sunday at eleven. Hence "the needle of the apparatus was up Sunday at eleven" is an event for Alice. But when does it become true? It becomes true Monday at noon when she hears Bob. That means that before Monday at noon, the needle position was indefinite. It becomes defined Monday at noon when Alice hangs-on to the corresponding branch. But immediately after, the event "the needle position was up Sunday at eleven" becomes true. So Monday at noon it becomes true that "the needle position was up Sunday at eleven" even if Monday at eleven thirty it was not true. One sees again that ConSol implies to change the way we consider the truth. The truth is relative to each observer's phenomenal world and, as such, applies only to propositions concerning events in this phenomenal world. The very fact that past events can emerge in this world only later and that before that they do not belong to it, shows that a proposition concerning the past can become true at a time T while it had no truth value at T-ε. This is precisely the reason why there is no backward in time effects. A backward in time effect is the fact that something happening at a time $T_1$ changes an event at a time $T_0 < T_1$. In the usual conception, the past is fixed and determined. So if something happening now was assumed to determine something supposed to have happened before, it would necessarily be a backward in time effect since what happened before (and is necessarily determined in the usual conception) has to change under the effect of the present action. But such effects do not exist. On the contrary, in ConSol, some past results are not necessarily definite (for example the value of the spin of a particle given by a measurement done before by another observer) and so, cannot be considered as events. What happens when a present measurement determine this value by choosing the corresponding branch is that the past result becomes definite and becomes an event. It is not that a past event has changed (because it was not definite). So it is very different from a backward in time effect.

Yes, that seems really weird but not so much if we remember that the reality relative to each observer (her phenomenal world) is composed by all the information contained in the branches the observer is hung-on to and only by that information. So it is not surprising that some allegedly past events are not determined. There is no absolute reality. So Alice's reality at time T contains all the events included in the branches Alice is hung-on to at time T. If we consider the entangled wave function of her empirical world at time T, Alice is hung-on to some of the branches corresponding to the measurements she did (and so are related to events), but many daughter branches still contain superpositions as in the example of wave function (10). For example, she measured the spin of Bob's particle because she talked to Bob Monday at noon but there are many other particles that she did not measure. The states of these particles can be entangled with the state of Bob's particle in many different ways and for many of them, their spin remains indefinite. Or they can be independent. For example, Alice can learn later, say Wednesday, that Charles measured the spin of one particle last week and found "+" and then it will become true Wednesday that the spin of this particle was "+" last week even though Tuesday it was indefinite for Alice.

The important point is that even though it seems that making true at a time T something that happened before seems to be a backward in time action, this is not the correct way to interpret the situation. A measurement is not something modifying the physical state of a system. It is just a random choice of one way to perceive an entangled universe (the empirical world that stays entangled). It is a construction of one's own phenomenal world from one's empirical world. A simplified (and not to take too seriously) way to picture that is the following: imagine that you watch a movie organized in such a way that each minute you must draw at random a document from an urn in order to decide what the next sequence is. The urn contains many documents each one telling stories consistent with what you saw before and with all the documents you have already drawn, but each one contains different suites. The document you draw at time T cannot contradict what you already know but can contain information about things having happened before but that were never mentioned previously. For example, at time T you draw a document saying that Paul married Mary a year ago. You did not know anything about that before. Hence, in the story you are watching, this becomes true. But it could have happened that at T, you had drawn a document saying that Paul died before getting married. A soon as nothing about Paul was mentioned in the previous documents you drew, this is consistent. Then what we notice is that before T, Paul was neither married nor dead. He was in an indefinite state. But immediately after T, it became true that Paul has been married a year ago with Mary even though five minutes before this was not true. The global entangled wave function is similar to the urn that contains all the possibilities[14]. Until you choose one, these possibilities remain virtual. Once you have drawn a document (you have made a measurement) what is implied by this document (by the result of the measurement) becomes true even if what is indicated happened in the past.

## 5. Conclusion

What we have shown is that no backward in time physical effect is necessary to explain the results of the various delayed choice experiments. Nevertheless, if we refuse to neglect the issues that the usual way to present the measurement raises inside a standard realist framework, explaining what happens in the case of the EPR experiment or in the delayed choice quantum eraser implies to find a new interpretation. This is what ConSol proposes. Then there is no more any contradiction and everything can be described in a coherent way. The price to pay is that we must renounce to the standard way to conceive the world and in particular, we must accept that the reality is not the same for all. We must also think differently of past events[15]. It can happen that a past property A of a system $S_1$ at $T_{-1}$ is not yet determined at a time $T_0 > T_{-1}$, but becomes determined when the observer makes at $T_1 > T_0 > T_{-1}$ a measurement of another property B (possibly on another system $S_2$) whose result is linked to the value that A should have had at $T_{-1}$ to be coherent with the result obtained on B at $T_1$. This is easy to understand in a framework where the reality of each observer is her own phenomenal world built through the results of the measurements that the observer performs in her empirical world. There is no physical effect from the future to the past but it can happen that some past results are indefinite in one observer's phenomenal world and become determined for this observer only after a measurement performed in their future. They become then events in the observer's phenomenal world.

---

[14] As we already noticed above, it is reminiscent of the consistent histories approach of Griffiths[23] and Omnes [24].

[15] Bitbol [26] made a similar assumption about the violation of the retrodiction principle to avoid non-locality but he stated it inside the many-worlds Everett's interpretation.